\documentclass[12pt]{article} 
\newcommand{\absosq}[1]{\bigl\vert#1\bigr\vert^2} 
\newcommand{\braket}[2]{\langle#1\vert#2\rangle} 
\newcommand{\ketbra}[2]{\vert#1\rangle\langle#2\vert} 
\newcommand{\ket}[1]{\vert#1\rangle} 
\newcommand{\bra}[1]{\langle#1\vert} 
\newcommand{\sandwich}[3]{\langle#1\vert#2\vert#3\rangle} 
\newcommand{\ts}[2]{\langle#1\Vert#2\rangle} 
\newcommand{\be}{\begin{equation}} 
\newcommand{\ee}{\end{equation}} 
\begin{document} 
\setlength{\baselineskip}{22pt} 
\title{Objective probabilities, quantum\\ 
counterfactuals, and the ABL rule} 
\author{Ulrich Mohrhoff\cite{ujm}\\ 
Sri Aurobindo International Centre of Education\\ 
Pondicherry-605002, India} 
\date{} 
\maketitle 
\begin{abstract} 
\noindent\normalsize The ABL rule is derived as a tool of standard quantum mechanics. The ontological 
significance of the existence of objective probabilities is discussed. Objections by Kastner [preceding 
article] and others to counterfactual uses of the ABL rule are refuted. Metaphysical presumptions 
leading to such views as Kastner is defending in her Comment are examined and shown to be 
unwarranted.\setlength{\baselineskip}{20pt} 
\end{abstract} 
 
\section{\large INTRODUCTION} 
 
Following Mermin~\cite{Mermin98}, I have characterized some of the probabilities that quantum 
mechanics allows us to calculate as being objective in the sense that they have nothing to do with 
ignorance---there is nothing for us to be ignorant of~\cite{Mohrhoff00}. I have argued that the 
objective probabilities associated with attributions of results to unperformed measurements should be 
calculated according to the ABL rule, first derived by Aharonov, Bergmann, and 
Lebowitz~\cite{ABL64}. Ruth Kastner~\cite{Kastner,Kastner99,Kastner99b} and 
others~\cite{SS93,Cohen95,Miller96} have raised objections concerning the appropriateness of 
probability assignments to counterfactuals based on the ABL rule. In the present article I re-derive the 
ABL rule, clarify the meaning of objective probabilities and the significance of their existence, and 
respond to these objections. 
 
In Sec.~2 both the Born rule and the ABL rule are derived from the standard, quantum-mechanical 
representation of contingent properties~\cite{noteCP} as projection operators on a Hilbert space. Section~3 refutes 
the objections by Kastner and the other authors to counterfactual uses of the ABL rule. Section~4 
elucidates the significance of the existence of objective probabilities, and Section~5 is devoted to 
unearthing unwarranted metaphysical presumptions leading to such views as Kastner is defending in 
her Comment~\cite{Kastner}. 
 
\section{\large OBJECTIVE PROBABILITIES AND THE ABL RULE} 
 
Quantum mechanics is unequivocal about the probabilities it assigns to the possible results of 
possible measurements. If we represent the potentially attributable~\cite{notePA} properties of a 
system as projection operators on a Hilbert space $\cal H$, there is no ambiguity as to the form of 
the prior probability measure $p\,(q_i,t)$, which is based solely on properties possessed by the 
system {\it before} the time $t$~\cite{Jauch68,Cassinello-SG96,Gleason57}: 
\be 
p\,(q_i,t)=\mbox{Tr}[{\bf W}(t){\bf P}_{q_i}]. 
\ee 
As is well known, $\bf W$ is a unique density operator [that is, a unique self-adjoint, positive operator 
satisfying $\mbox{Tr}({\bf W})=1$ and ${\bf W}^2\leq{\bf W}$]. $\mbox{Tr}$ signifies the trace defined 
by the formula $\mbox{Tr}({\bf X}):=\sum_i\sandwich{i}{{\bf X}}{i}$ for any orthonormal basis 
$\{\ket{i}\}$ in $\cal H$. If ${\bf W}^2={\bf W}$, ${\bf W}(t)$ projects on a one-dimensional subspace 
of $\cal H$ and thus is equivalent---apart from an irrelevant phase factor---to a ``state'' vector 
$\ket{\psi(t)}$. Such a ``state'' is said to be ``pure,'' and the system is said to be ``prepared'' in it. With 
${\bf W}(t)=\ketbra{\psi(t)}{\psi(t)}$ and ${\bf P}_{q_i}=\ketbra{q_i}{q_i}$, we obtain the familiar Born 
rule: 
\be 
\label{Born} 
p\,(q_i,t)=\mbox{Tr}[\ketbra{\psi(t)}{\psi(t)}{\bf P}_{q_i}]
=\sandwich{\psi(t)}{{\bf P}_{q_i}}{\psi(t)}=\absosq{\braket{\psi(t)}{q_i}}. 
\ee 
Applying the Born rule twice, we obtain the prior probability that a system with a prior 
probability measure $\ketbra aa$ will first be observed to have the property $\ketbra{q_i}{q_i}$ and 
then be found in possession of the property $\ketbra bb$: 
\be 
p\,(q_i,b|a)=\absosq{\braket a{q_i}\braket{q_i}b}. 
\ee 
(For simplicity's sake we will assume that the Hamiltonian is zero between measurements.) Although  
readers familiar  with the mathematical formalism of quantum mechanics are not likely to stumble 
over this  expression, it involves a conceptual transition that needs to be justified. When we ask for 
the  probability that $q_i$, given $a$, the projection operator $\ketbra{q_i}{q_i}$ represents a 
potentially attributable property of the system. When we ask for the probability that $b$, given $q_i$, 
the same operator represents a probability measure. How do we get from a property to a probability 
measure?

If we start out with a probability measure ${\bf W}_1=\ketbra aa$ and then find that the system 
has the property $\bf P$, we must update our probability measure accordingly. If the 
measurement that yields the property $\bf P$ is ideal---as is generally assumed when discussing 
interpretational issues,---the updated probability measure assigns probability zero to any property 
${\bf P}'$ for which ${\bf P}{\bf P}'=0$. Hence the ``state'' of the system ``collapses''---not 
mysteriously but self-evidently---to the ``state'' 
\be 
{\bf W}_2={{\bf P}\ketbra aa{\bf P}\over\sandwich a{{\bf P}}a}. 
\ee 
The denominator ensures that the probability of the trivial property, represented by the identity 
operator $\bf 1$, remains~1. If we put $\ketbra{q_i}{q_i}$ in place of $\bf P$, this reduces to 
${\bf W}_2=\ketbra{q_i}{q_i}$. Thus the updated probability measure is represented by the same 
operator as the property observed~\cite{note1}. 
 
Next, we consider the probability that a system with a prior probability measure $\ketbra aa$ will 
be found in possession of property $\ketbra bb$ given that in the meantime $Q$ is measured but 
regardless of the result of this measurement. This is obviously the sum of probabilities 
\be 
p\,(b|a,Q)=\sum_j\absosq{\braket a{q_j}\braket{q_j}b}. 
\ee 
According to Bayes' theorem, the probability that the intervening measurement yields $q_i$ given that 
the prior probability measure is $\ketbra aa$ and given that the final measurement yields $b$, 
then is 
\be 
\label{ABL} 
p\,(q_i|a,b)={p\,(q_i,b|a)\over p\,(b|a,Q)}= 
{\absosq{\braket a{q_i}\braket{q_i}b}\over\sum_j\absosq{\braket a{q_j}\braket{q_j}b}}. 
\ee 
This is (one of the possible forms of) the ABL rule. Like the Born rule, it follows straight from the 
quantum-mechanical representation of contingent properties as projection operators on 
a Hilbert space. 
 
In principle, both rules have an objective as well as a subjective application. If $Q$ is actually 
measured, both rules assign probabilities that are subjective inasmuch as they are based on probability 
measures that fail to take account of at least one relevant fact---the result of the measurement of 
$Q$. In order to be considered objective, a quantum-mechanical probability must be assigned on the 
basis of all relevant facts. If the $Q$ measurement is actually made, the objective probabilities 
associated with its possible outcomes are trivially either zero or one. Thus both rules can assign 
nontrivial objective probabilities only if the $Q$ measurement is not made. But this is not sufficient. 
Since Born probabilities take no account of (future) facts about the system's future properties, they can be 
considered objective only if there are no relevant such facts. This is hardly ever the case. Hence, in 
general, objective probabilities are calculated according to the ABL rule. We can drop the qualifying 
``in general'' if we use the trivial property in place of $\ketbra bb$ if there never will be any facts about the 
system's future properties. In this case the ABL rule reduces to the Born rule: 
\be 
{\absosq{\braket a{q_i}}\sandwich{q_i}{{\bf1}}{q_i}\over 
\sum_j\braket a{q_j}\braket{q_j}a\sandwich{q_j}{{\bf1}}{q_j}} 
={\absosq{\braket a{q_i}}\over\bra a\left(\sum_j\ketbra{q_j}{q_j}\right)\ket a} 
=\absosq{\braket a{q_i}}. 
\ee 
Thus objective probabilities are calculated according to the ABL rule, and they are assigned to 
contrary-to-fact conditionals, or counterfactuals, of the following general form: 
\begin{description} 
\item[(A)] If a measurement of observable $Q$ were performed on system $S$ between an actual 
measurement yielding the result $\ketbra aa$ at time $t_a$ and an actual measurement yielding the 
result $\ketbra bb$ at time $t_b$, but no measurement is actually performed between $t_a$ and 
$t_b$, then the measurement of $Q$ would yield $q_i$ with probability $p\,(q_i|a,b)$. 
\end{description} 
For $p\,(q_i|a,b)$ to be objective, it is not enough that $Q$ is not actually measured; it is necessary 
that {\it no} measurement is performed between $t_a$ and $t_b$. If any other measurement $M$ is 
performed during this time span, $p\,(q_i|a,b)$ is based on an incomplete set of facts---it does not 
take account of the result of $M$---and is therefore subjective. Note that the antecedent clause states not  
only that the $Q$ measurement is made (counterfactually or in a possible world) but also that at the  
times $t_a$ and $t_b$ the respective results $\ketbra aa$ and $\ketbra bb$ are obtained (that is, they  
are as in the actual world). The observation that the result obtained at $t_b$ might have been different had the 
$Q$ measurement been actually made, is irrelevant to the truth of (A) inasmuch as it is based on an 
incomplete set of facts---it does not take into account the result obtained at $t_b$---while statement 
(A) is based on a complete set of facts, as it must be in order to assign objective probabilities. 
Probabilities are objective only if they are assigned to unperformed measurements, and only if the 
assignment is based on all relevant facts. The second ``if'' translates into the strong {\it ceteris 
paribus} clause that all measurement results obtained in the actual world are also obtained in the 
possible worlds considered by~(A). These possible worlds differ from the actual world only in that they 
contain one extra measurement that is not made in the actual world.

\section{\large REPLY TO KASTNER} 
 
Ruth Kastner~\cite{Kastner,Kastner99,Kastner99b} and others~\cite{SS93,Cohen95,Miller96} have 
raised objections concerning the appropriateness of probability assignments to counterfactuals based 
on the ABL rule. The first objection I will address is this~\cite{Kastner99b}: Since the $Q$ 
measurement is not actually made, the following rule should be used instead of the ABL rule 
(\ref{ABL}): 
\be 
\label{Kastner} 
p_K(q_i|a,b)={p\,(q_i,b|a)\over p\,(b|a)} 
={\absosq{\braket a{q_i}\braket{q_i}b}\over\absosq{\braket ab}} 
={\absosq{\braket a{q_i}\braket{q_i}b}\over\absosq{\sum_j\braket a{q_j}\braket{q_j}b}}. 
\ee 
This rule combines, according to Bayes' theorem, the probability $p\,(q_i,b|a)$ 
that a system with a prior probability measure $\ketbra aa$ will first be observed to have 
property $\ketbra{q_i}{q_i}$ and then be found in possession of property $\ketbra bb$, with the 
probability $p\,(b|a)$ that an equally ``prepared''~\cite{noteB} system will be observed to have 
property $\ketbra bb$ given that in the meantime no measurement is made. In the numerator 
Kastner assumes that $Q$ is measured, and in the denominator she assumes that between $t_a$ and 
$t_b$ no measurement takes place. Whereas in the denominator of the ABL rule (\ref{ABL}) 
probabilities are added---the $Q$ measurement is assumed to be made,---in the denominator of 
Kastner's rule (\ref{Kastner}) amplitudes are added, which entails that the $Q$ measurement is not 
made. The ABL rule thus is consistent---it assumes throughout that the $Q$ measurement is made,--- 
while the same cannot be said of Kastner's rule. 
 
A similar inconsistency mars arguments by Sharp and Shanks~\cite{SS93}, Cohen~\cite{Cohen95}, 
Miller~\cite{Miller96}, and Kastner~\cite{Kastner99,Kastner99b} purporting to prove the general 
invalidity of counterfactual uses of the ABL rule. These arguments have been refuted---cogently, in 
my opinion---by Vaidman~\cite{Vaidman99}, though Kastner~\cite{Kastner99a}, predictably, takes a 
different view. What these ``proofs'' purport to show is that the counterfactual use of the ABL rule 
yields results that are inconsistent with standard quantum mechanics. Specifically, it is claimed that 
this use entails the following equation: 
\begin{eqnarray} 
\label{BS} 
p\,(q_j|a)&=&p\,(q_j|a,b_1)\, p\,(b_1|a)+ p\,(q_j|a,b_2)\, p\,(b_2|a)\nonumber\\ 
&=&{p\,(q_j,b_1|a)\over\, p\,(b_1|a,Q)}\,p\,(b_1|a) 
+{p\,(q_j,b_2|a)\over\, p\,(b_2|a,Q)}\,p\,(b_2|a). 
\end{eqnarray} 
Since the final measurement of the observable $B$, assumed to have two eigenvalues 
$b_1$ and $b_2$, is actually made, the Born probability $p\,(q_j|a)$ of the outcome $q_j$ of an 
intermediate measurement of $Q$ is uncontroversially the sum of the probabilities $p\,(q_j,b_1|a)$ 
and $p\,(q_j,b_2|a)$. According to eq. (\ref{BS}) this is not always the case. It is the case if 
\begin{eqnarray} 
p\,(b_i|a,Q)&=&\sum_j\absosq{\braket a{q_j}\braket{q_j}{b_i}}\nonumber\\ 
&=&\left|\sum_j\braket a{q_j}\braket{q_j}{b_i}\right|^2=p\,(b_i|a). 
\end{eqnarray} 
This holds if $Q=A$ (the observable measured at time $t_a$) or $Q=B$, or if for $j\neq k$, 
\be 
\Re\left(\braket{q_k}a\braket a{q_j}\braket{q_j}{b_i}\braket{b_i}{q_k}\right)=0. 
\ee 
Kastner states these conditions in Refs. \cite{Kastner99} and~\cite{Kastner99b}. It is then argued that  
since the counterfactual use of the ABL rule in eq. (\ref{BS}) generally leads to inconsistencies  with 
standard quantum mechanics, this use is illegitimate unless one of  those conditions is satisfied. 
However, what is illegitimate is not the counterfactual use of the ABL  rule but the equation on which 
this conclusion is based. Like Kastner's rule (\ref{Kastner}), eq. (\ref{BS}) combines expressions that 
imply that the intervening measurement is made---namely,  $p\,(q_j|a,b_1)$ and 
$p\,(q_j|a,b_2)$---with expressions that imply the contrary, namely, $p\,(b_1|a)$  and $p\,(b_2|a)$. 
Note that it would not improve matters if instead of the ABL probabilities  $p\,(q_j|a,b_i)$ the 
probabilities $p_K(q_j|a,b_i)$ were used in eq. (\ref{BS}) [which would  ensure that 
$p\,(q_j|a)=p\,(q_j,b_1|a)+p\,(q_j,b_2|a)$] since the probabilities $p_K(q_j|a,b_i)$ involve the same 
inconsistency. In order to be consistent we must assume throughout that the intervening 
measurement  is made. This entails that instead of the probabilities $p\,(b_i|a)$ the probabilities 
$p\,(b_i|a,Q)$ must  be used in eq. (\ref{BS}), which likewise ensures that 
$p\,(q_j|a)=p\,(q_j,b_1|a)+p\,(q_j,b_2|a)$. 
 
One might perceive a contradiction between the counterfactual (A) and my insistence on the 
necessity of assuming consistently that the intervening measurement is made. But this apparent 
``contradiction'' is the very nature of a counterfactual. Counterfactuals are statements about 
conceivable worlds that are different from the actual world in certain specified respects and like the 
actual world in all other relevant respects. The worlds considered by (A) are different from the actual 
world in that between $t_a$ and $t_b$ a measurement is made that is not made in the actual world; in 
all other respects they are like the actual world. In particular, the measurements at $t_a$ and $t_b$ 
have the specified outcomes. Since (A) is a statement about conceivable worlds in which the $Q$ 
measurement is made, the probabilities it assigns to the possible outcomes of this measurement must 
have the same values as the probabilities that we would assign, on the same basis, to the possible 
outcomes of the same measurement, if this were actually made. 
 
Kastner denies the validity of the counterfactual (A) also on the ground that it allegedly violates the 
requirement of cotenability. A counterfactual $C$ refers to conceivable worlds that must satisfy two 
conditions:  They must be different from the actual world in certain specified respects, and they must 
be like the  actual world in all other relevant respects. If it turned out that there are no conceivable 
worlds that satisfy both conditions, $C$ would fail to satisfy this crucial requirement.
 
Consider a conceivable world ${\cal W}_i$ in which the $Q$ measurement yields an 
outcome $q_i$ to which (A) assigns probability zero. (To each possible outcome $q_k$ there 
corresponds a conceivable world ${\cal W}_k$.) ${\cal W}_i$ fails to satisfy the second condition 
inasmuch as it is unlike the actual world in another relevant respect: It does not obey the actual 
physical laws. But this is something that we learn from (A) rather than something that invalidates~(A). 
Statement (A) itself tells us that the conceivable world ${\cal W}_i$ is not a nomologically possible 
world inasmuch as under the specified conditions a measurement of $Q$ would never yield the result $q_i$. 
 
What Kastner has in mind is something more serious. According to the metalinguistic account of 
counterfactuals~\cite{Goodman47}, invoked by Kastner~\cite{Kastner99}, a counterfactual is true if its 
{\it antecedent} conjoined with {\it laws of nature} and statements of {\it background conditions} 
logically entails its consequent. The background conditions must be stable (that is, they 
must hold independently of the truth value of the antecedent). Kastner claims that (A) violates either 
the requirement of cotenability---the background conditions depend on the truth value of the 
antecedent---or the laws of nature.

Before addressing Kastner's argument I will show that both the stability of the background conditions 
and the laws of nature are in fact respected by (A). The antecedent is the statement that a 
measurement of observable $Q$ is performed 
on system $S$ between $t_a$ and $t_b$. The background conditions consist in the observation (or 
the factually warranted possession) of the properties $\ketbra aa$ and $\ketbra bb$ at the respective 
times $t_a$ and $t_b$, as well as in the absence of any measurement between $t_a$ and $t_b$ other 
than that specified in the antecedent. The relevant laws of nature are the principles of quantum 
mechanics. Being statistical laws, they enable us to assign probabilities to the possible results of 
measurements, and being universal laws that have never been found to conflict with experimental 
data, they allow us to apply them counterfactually---to assign probabilities to the possible results of 
unperformed measurements. These assignments can be made on the basis of all relevant facts about 
either the past or the future properties of $S$ using the Born rule, or on the basis of all relevant facts 
about the past and future properties of $S$ using the ABL rule. Since both rules are part of standard 
quantum mechanics, none of these assignments can conflict with standard quantum mechanics. Nor can they 
conflict with the required stability of background conditions. If obtaining $\ketbra aa$ at $t_a$ and 
$\ketbra bb$ at $t_b$ is nomologically possible without interposition of any measurement, the same is 
nomologically possible whenever a measurement is interposed~\cite{noteIP}. The reason this is so 
is that the interposition of a measurement never decreases the probability of obtaining $\ketbra bb$ at 
$t_b$ given $\ketbra aa$ at $t_a$: 
\be 
p\,(b|a)=\absosq{\braket ab} = \absosq{\sum_k \braket a{q_k}\braket{q_k}b} 
\leq\sum_k \absosq{\braket a{q_k}\braket{q_k}b }=p\,(b|a,Q). 
\ee 
Thus the background conditions are consistent with both the truth and the falsity of 
the antecedent. And given the necessity of assuming consistently that the intervening measurement 
is made, the antecedent, conjoined with the relevant laws of nature and the background condition 
statement, logically entails that the probability of obtaining the value $q_i$ is as given by the ABL 
rule~\cite{notePC}. 
 
If Kastner reaches a different conclusion, it is because her understanding of the background 
conditions and/or of the relevant laws of nature is different from mine. That she thinks differently 
about the background conditions is obvious from her discussion~\cite{Kastner99} of the experiment 
considered by Sharp and Shanks~\cite{SS93}. In this experiment spin-$1\over2$ particles are 
prepared at time $t_1$ with probability measure $\ket{a_+}$ and subjected at time $t_2$ to a 
measurement of their spin component along the $\bf b$ axis. We are accustomed to read 
$\ket{a_+}$ as ``spin up along direction $\bf a$'', but what $\ket{a_+}$ really signifies depends on the 
time to which it refers. While at $t_1$ the system possesses the property represented by 
$\ketbra{a_+}{a_+}$, for $t>t_1$ the ket $\ket{a_+}$ or the density operator ${\bf 
W}=\ketbra{a_+}{a_+}$ represents a probability measure that says nothing about properties 
possessed at $t$; it only tells us that the prior probability of obtaining the result ``spin up along 
direction $\bf c$'' at the time $t$ is $\sandwich{c_+}{{\bf W}}{c_+}$. By the same token, if the final 
measurement indicates that the property represented by the operator $\ketbra{b_+}{b_+}$ is 
possessed at $t_2$, then this operator also represents the ``time-reversed'' density operator ${\bf 
W}'$ that yields the posterior probability~\cite{notePP} $\sandwich{c_+}{{\bf W}'}{c_+}$ of obtaining 
the result ``spin up along direction $\bf c$'' at the time~$t<t_2$. 
 
Kastner, following Sharp and Shanks, states that the measurement at $t_2$ yields a mixture $M$ 
consisting of two subensembles. Thus she considers an ensemble of measurements, performed on an 
ensemble of systems, rather than an individual measurement, for an individual measurement does 
not yield a {\it mixture} consisting of subensembles; it yields a {\it result}, in this case either the 
property represented by $\ketbra{b_+}{b_+}$ or the property represented by $\ketbra{b_-}{b_-}$. 
Thereafter  Kastner states what she takes to be ``the basic conceptual problem'': ``in considering a 
counterfactual  measurement of the spin along $\bf c$ (observable $\sigma_c$) [at an intermediate 
time $t$] we  must take into account all the effects of that measurement on the system. The 
measurement of the  observable $\sigma_c$ results in a change in the mixture $M$ of post-selected 
ensembles'' into a  different mixture $M'$. Kastner goes on to say that ``the mixture, $M$ or $M'$, 
obtaining at time  $t_2$ \ldots must enter into the counterfactual calculation,'' and that ``the 
characterization of the  mixture obtaining at $t_2$ must be included in the background condition 
statement~\ldots'' Since  this mixture depends on whether or not the $\sigma_c$ measurement is 
performed at the time $t$,  Kastner concludes that the statement of background conditions holding 
when the antecedent is false  (no intervening measurement) becomes false when the antecedent 
holds, and is therefore not  cotenable with the antecedent. 
 
Kastner is led to this fallacious conclusion by conflating statements about ensembles with 
statements about individual systems. It is true that if we start with an ensemble of systems possessing 
property $\ketbra {a_+}{a_+}$ at time $t_1$, the result or effect of an ensemble of $\sigma_b$ 
measurements performed at $t_2$ is a mixture $M$ consisting of two subensembles, one containing 
systems possessing the property $\ketbra{b_+}{b_+}$, and another containing systems possessing 
the property $\ketbra{b_-}{b_-}$, while the result or effect of two ensembles of measurements, one of 
$\sigma_c$ performed at time $t$ and one of $\sigma_b$ performed at time $t_2$, is a mixture $M'$  
consisting  of four subensembles corresponding to that many combinations of possible measurement 
outcomes.  But all this is irrelevant to the truth of (A) or the cotenability of its antecedent with its 
background  conditions, for (A) is a statement about an individual system, not a statement about an 
ensemble. The  {\it only} effect of the intervening measurement on an individual system, under the 
specified  background conditions, is that at the time $t$ it has either the property $\ketbra{c_+}{c_+}$ 
or the  property $\ketbra{c_-}{c_-}$, neither of which it has if the intervening measurement is not 
made. The  relevant question is not: Might the final measurement have a different outcome from the 
one it  actually has if the $\sigma_c$ measurement were performed? The relevant question is: Are 
the  background conditions---$\ketbra{a_+}{a_+}$ possessed at $t_1$ and $\ketbra{b_+}{b_+}$ 
possessed  at $t_2$---consistent with both the truth and the falsity of the antecedent? The answer is 
affirmative, and this is sufficient for cotenability and hence for the legitimacy of (A). 
 
In her Comment~\cite{Kastner}, Kastner introduces one Dr.~X who asks himself the irrelevant 
question: ``How might the data of my experiment have changed if I had made a measurement at time 
$t$ that I did not, in fact, make?'' Obviously, Dr.~X might not have obtained the result 
$\ketbra bb$ at the time $t_b$. The counterfactual (A) addresses a different question in that it rules out this 
possibility. While Dr.~X's question assumes the possession of property $\ketbra aa$ at time $t_a$, (A) 
in addition assumes the possession of property $\ketbra bb$ at time $t_b$. According to 
Kastner, one gets from the question asked by Dr.~X to the question addressed by (A) by requiring the 
following: ``If a measurement of observable $Q$ had been performed, system $S$ would (with 
certainty) have been pre- and post-selected with outcomes $a$ and $b$ as in the actual world.'' This 
requirement, so Kastner claims, correctly expresses the cotenability that is necessary for the 
consistency and the truth of (A). And since this requirement is obviously ``not guaranteed to hold,'' 
she concludes that the counterfactual statement (A) fails~\cite{noteNC}. 

By requiring that system $S$ would {\it with certainty} have the properties $a$ and $b$ at the 
respective time $t_a$ and $t_b$, Kastner introduces an element of nomological necessity that makes 
nonsense of (A). What she thereby refutes is not (A) but the following counterfactual (A$'$), which 
requires no refutation because it is patently false: 
\begin{description} 
\item[(A$'$)] If a measurement of observable $Q$ were performed on an ensemble of systems 
between an actually performed measurement of observable $A$ at time $t_a$ and an actually 
performed measurement of observable $B$ at time $t_b$, but no measurement is actually made 
between $t_a$ and $t_b$, then the measurement of $Q$ would yield $q_i$ with probability 
$p\,(q_i|a,b)$, and the measurements of $A$ and $B$ would (with certainty) yield the respective 
outcomes $a$ and $b$.'' 
\end{description} 
The counterfactual (A) attributes a probability to a possible result of a possible measurement 
hypothetically performed at time $t$ on an {\it individual system} $S$ which (in the actual world) {\it 
happens} to possess the properties $a$ and $b$ at the respective times $t_a$ and $t_b$. As has 
been explained at length in Ref.~\cite{Mohrhoff00}, nothing ever causes (i)~a measurement, {\it qua} 
attempt to determine the value of some observable, to yield a result or (ii)~a measurement, {\it qua} 
successful determination of the value of some observable, to take place. In other words, the actual 
events or states of affairs that indicate the possession of a contingent property (by a 
system) or of a value (by an observable) are causal primaries, and this not in the sense that nothing 
ever causes a measurement to yield this particular value rather than that, but in the sense that 
nothing ever causes a measurement to be successful or to take place. (A causal primary is an event 
or state of affairs the occurrence or existence of which is not necessitated by any cause, antecedent 
or otherwise.) What the laws of quantum mechanics encapsulate is statistical correlations between 
causal primaries. If we take certain measurement results as given, we can use those laws to assign 
probabilities to the possible results of other measurements, which may or may not be performed. In 
particular, we can consider the possession of $a$ (at time $t_a$) and of $b$ (at time $t_b$) as given, 
and use the laws of quantum mechanics to assign probabilities to the possible results of a not actually 
performed  intervening measurement. In so doing we do {\it not} assume that the occurrences of the  
measurements at $t_a$ and $t_b$ are necessitated by anything, let alone that the respective  
outcomes $a$ and $b$ are. 
 
The only way to test probability assignments is to determine relative frequencies with the help of 
appropriately selected ensembles. In the case of (A), the appropriate ensemble is an ensemble of 
possible worlds in which system $S$ is identically ``prepared'' or ``pre-selected'' as well as identically 
``retropared'' or ``post-selected,'' and in which the $Q$ measurement is made. Although Kastner 
introduces one Dr.~X$^\dagger$ who is ``associated with'' all of those possible worlds, it is obvious 
that this ensemble is not empirically accessible. It is possible, however, to reproduce this ensemble in 
the actual world, by turning it into an ensemble of identically ``prepared'' and ``retropared'' copies of 
$S$. In order to render testable the probabilities that (A) assigns, we have no choice but to use an 
ensemble of pre- and post-selected systems. Yet the probabilities that (A) assigns are single-case 
probabilities; they are assigned to the possible outcomes of a single measurement on a single system 
that {\it happens} to possess property $a$ at time $t_a$ and property $b$ at time $t_b$. The 
selection does not reflect any nomological constraint but merely serves to make those probabilities 
measurable. 
 
Kastner is committed to denying that counterfactual probability assignments can be tested inasmuch 
as she considers the probabilities that can be measured (using pre- and post-selected ensembles) to 
differ quantitatively from the corresponding counterfactual probabilities. Accordingly, she considers 
statement (A) to be distinct from the following statement: 
\begin{description} 
\item[(B)] In a possible world in which observable $Q$ is measured at time $t$ and 
system $S$ yields outcomes $\ketbra aa$ and $\ketbra bb$ at times $t_a$ and $t_b$, respectively, 
the probability of obtaining result $q_i$ is given by $p\,(q_i|a,b)$. 
\end{description} 
To claim a quantitative difference between counterfactual and non-counterfactual assignments of 
quantum-mechanical probabilities or a significant difference between statements (A) and (B) is to 
misunderstand the meaning of counterfactual assignments of quantum-mechanical probabilities. 
Saying that $p\,(q_i|a,b)$ is the (objective) probability with which $q_i$ {\it would be} obtained, given 
the outcomes $\ketbra aa$ and $\ketbra bb$ at the respective times $t_a$ and~$t_b$, is in all relevant 
respects exactly the same as saying that $p\,(q_i|a,b)$ is the (subjective) probability with which $q_i$ 
{\it is} obtained given the same outcomes at times $t_a$ and~$t_b$. What else could statement (A) 
possibly mean? 
 
None of the arguments marshaled by Kastner and the authors cited by her succeed in proving 
that statement (B) ``differs significantly'' from statement~(A). That the outcome at $t_b$ might be 
different from $\ketbra bb$ if the $Q$ measurement were made, given the outcome at $t_a$ alone, is 
irrelevant since the outcomes at $t_a$ and $t_b$ are both given. The argument from cotenability fails 
because it conflates statements about individual systems with statements about ensembles of 
systems. The attempt to replace $p\,(q_i|a,b)$ (eq.~\ref{ABL}) by $p_K(q_i|a,b)$ 
(eq.~\ref{Kastner}) in (A) and the attempt to show that counterfactual uses of the ABL rule yield 
consequences that are inconsistent with quantum theory, both fail because they combine expressions 
that imply that the $Q$ measurement is both performed and not performed. One does not do justice 
to the counterfactuality of probability assignments by arguments that begin by assuming that $Q$ is 
not measured and end up by assuming that $Q$ is measured. One does it justice by considering 
possible worlds in which $Q$ is measured or a suitably pre- and post-selected ensemble of 
actual-world systems. 
 
\section{\large THE SIGNIFICANCE OF OBJECTIVE\\
PROBABILITIES} 
 
In this section I want to elucidate the significance of the existence of objective probabilities. Such 
probabilities are not merely best guesses based on a complete knowledge, and thus free from any 
element of ignorance; they also tell us something important about the objective world. 
 
To begin with, let us consider the following experiment~\cite{Vaidman96}. A particle initially in 
possession of the 
property $\ketbra{\psi_1}{\psi_1}$ is eventually found in possession of the property 
$\ketbra{\psi_2}{\psi_2}$, where $\ket{\psi_1}=(\ket A+\ket B+\ket C)/\sqrt3$ and $\ket{\psi_2}=(\ket 
A+\ket B-\ket C)/\sqrt3$. The projection operators ${\bf P}_A=\ketbra AA$, ${\bf P}_B$, and ${\bf 
P}_C$ represent the respective properties of being inside one of three sealed boxes $A$, $B$, and 
$C$ at an intermediate time $t$. The ABL probability of finding the particle inside box $A$ at the 
time $t$ is 
\be 
\label{boxes} 
p\,(A|\psi_1,\psi_2)={\absosq{\sandwich{\psi_1}{{\bf P}_A}{\psi_2}}\over 
\sum_j\absosq{\sandwich{\psi_1}{{\bf P}_j}{\psi_2}}}. 
\ee 
As it stands, $p\,(A|\psi_1,\psi_2)$ is underdetermined. To assign to it a value, we still have 
to specify exactly which observable is being measured. If it is the observable $Q$ whose 
eigenkets are $\ket A$, $\ket B$, and $\ket C$, the denominator is given by 
\be 
\label{deno2} 
\absosq{\sandwich{\psi_1}{{\bf P}_A}{\psi_2}}+\absosq{\sandwich{\psi_1}{{\bf P}_B}{\psi_2}} 
+\absosq{\sandwich{\psi_1}{{\bf P}_C}{\psi_2}}={1\over9}+{1\over9}+{1\over9}={1\over3}. 
\ee 
Since the numerator is equal to $1/9$, $p\,(A|\psi_1,\psi_2)$ is equal to $1/3$. 
If on the other hand we measure the binary observable $Q_A={\bf P}_A$, the denominator is given by 
\be 
\label{deno1} 
\absosq{\sandwich{\psi_1}{{\bf P}_A}{\psi_2}}+\absosq{\sandwich{\psi_1}{({\bf P}_B+{\bf 
P}_C)}{\psi_2}} ={1\over9}+0={1\over9}, 
\ee 
and $p\,(A|\psi_1,\psi_2)$ is equal to $1$. By the same token, if we measure $Q$ then 
$p\,(B|\psi_1,\psi_2)=1/3$, and if we measure $Q_B$ then $p\,(B|\psi_1,\psi_2)=1$. 
 
I would like to discuss these results in the context of a somewhat more realistic setup. Consider a wall 
$W$ in which there are three holes $A$, $B$ and $C$. In front of the wall there is a particle source 
$E$. Behind the wall there is a particle detector $D$. Both $E$ and $D$ are equidistant from the 
three holes. Behind $C$ there is a device that causes a phase shift by $\pi$. ${\bf P}_j$ now 
represents the alternative ``the particle goes through hole $j$'', where $j$ may also stand for a union 
like $B\cup 
C$, the opening made up of $B$ and $C$. For particles emitted by $E$ the prior probability measure 
(with respect to the time at which they pass the wall) thus is $\ket{\psi_1}$, and for particles detected 
by $D$ the posterior probability measure (with respect to the same time) is $\ket{\psi_2}$. To 
measure $Q_A$, we place near $A$ a device $F_A$ that beeps whenever a particle passes through 
$A$. To measure $Q_B$, we place near $B$ a device $F_B$ that beeps whenever a particle passes 
through $B$. To measure $Q$, we use both devices. (If the devices are 100\% efficient, the absence 
of a beep then tells us that the particle went through $C$.) 
 
What we just found is this: If only $F_A$ is in place, the particle goes through $A$ with probability 
one (assuming, of course, that it is both emitted by $E$ and detected by $D$). If only $F_B$ is in 
place, the particle goes through $B$ with probability one. If both beepers are in place, the particle is 
equally likely to go through any of the three holes. Hence ABL probabilities in general are {\it 
contextual}---they depend on the distinctions that a particular setup permits us to 
make~\cite{noteC}. By 
measuring $Q$ we can tell whether the particle goes through $A$, through $B$, or through $C$. By 
measuring $Q_A$, we can tell whether it goes through $A$ or through $B\cup C$. With the former 
setup (both beepers in place) three properties are available for predication---``through $A$'', 
``through $B$'', and ``through $C$'',---with the latter (one beeper in place) only two are available. With 
the former setup three spatial distinctions are warranted---between $A$ and $B$, between $A$ and 
$C$, and between $B$ and $C$,---with the latter only one is warranted. $p\,(A|\psi_1,\psi_2)$ 
depends on the spatial distinctions that are warranted within the complement of $A$ in $A\cup B\cup 
C$. If none are warranted by the experimental setup then $p\,(A|\psi_1,\psi_2)=1$. If the distinction 
between $B$ and $C$ is warranted by the setup then $p\,(A|\psi_1,\psi_2)=1/3$. 
 
The contextuality of objective ABL probabilities makes it obvious that probability one does not imply 
actual possession of a value (by an observable) or of a property (by a physical system): From our 
ability to infer with probability one the result of measuring a physical quantity at time~$t$, it does not 
follow that at the time $t$ there exists an element of reality corresponding to the physical quantity 
and having a value equal to the predicted measurement result. Hence the {\it only} reason we have 
for attributing a contingent property $q$ to a physical system or a value $v$ to a 
quantum-mechanical observable is the occurrence/existence of an actual event or state of affairs 
from which the possession of $q$ or $v$ can be inferred, or by which it is indicated. In other words, 
the contextuality of ABL probabilities implies that the contingent properties of physical systems are {\it 
extrinsic}: They supervene on what happens or is the case in the rest of the world. To paraphrase 
Wheeler~\cite{Wheeler}, {\it no property is a possessed property unless it is an indicated property}. In 
short, owing to their contextuality, ABL probabilities cannot be assigned without specifying the range 
of values of an observable that is measured or assumed to be measured; and owing to their extrinsic 
nature, contingent properties cannot be attributed unless their possession is warranted by 
facts~\cite{noteD}. 
 
It may be held that, unlike an ABL probability equal to one, a Born probability equal to one is 
sufficient for the possession of a value~\cite{Redhead87}, or that an element of reality corresponding 
to an eigenvalue of an observable $Q$ exists at a time $t$ if the Born probability measure for the 
time $t$ has the pure form $\ketbra{\psi(t)}{\psi(t)}$ and $\ket{\psi(t)}$ is an eigenstate of $Q$. But 
this is an error. The extrinsic nature of contingent properties can be established without invoking ABL 
probabilities~\cite{Mohrhoff00}. The contextuality of ABL probabilities merely confirms 
it~\cite{noteCT}. 

Cohen~\cite{Cohen95} states that in standard quantum mechanics, probabilities for obtaining 
particular results are not contextual. What he means is that {\it Born} probabilities are not contextual. 
As was shown in Sec.~2, both the Born rule and the ABL rule follow straight from the 
quantum-mechanical representation of contingent properties as projection operators on a Hilbert 
space. Both 
therefore are tools of standard quantum mechanics. Cohen further states that the product rule is 
always valid in standard quantum mechanics. According to the product rule, if $X$ and $Y$ are 
commuting observables, if a measurement of $X$ will yield $x$ with Born probability one and a 
measurement of $Y$ will yield $y$ with Born probability one, then a measurement of $XY$ will yield 
$xy$ with Born probability one. This too is a statement about Born probabilities, not a statement about 
standard quantum mechanics. ABL probabilities do not always satisfy the product rule, and therefore 
standard quantum mechanics does not always satisfy the product rule. The operators $Q_A={\bf 
P}_A$ and $Q_B={\bf P}_B$ commute; an intervening measurement of $Q_A$ would yield $1$ 
(``through $A$'') with probability one; an intervening measurement of $Q_B$ would yield 1 (``through 
$B$'') with probability one; yet a measurement of $Q_AQ_B$ would yield nothing because 
(i)~$Q_AQ_B=0$ and (ii)~there is no world in which $Q_AQ_B$ can be measured. While $Q_A$ is 
measured in worlds in which {\it only} $F_A$ is in place, $Q_B$ is measured in worlds in which {\it 
only} $F_B$ is in place. Hence it is {\it logically} impossible for both $Q_A$ and $Q_B$ to be 
measured in the same world. What is measured in a world in which both beepers are in place is not 
$Q_AQ_B$ but $Q$. Thus the product rule is ``violated'' only by combinations of counterfactual 
statements referring to different possible worlds. 
 
Either type of probability has its specific use. The ABL rule is obviously of no use for {\it predicting}, on the basis of data obtained before time $t$,
the result of a measurement performed at the time $t$, for its application presupposes knowledge of 
the results of measurements performed after the time $t$. [Even if the measurement at the time $t$ is the 
last measurement ever to be performed on the system, we would have to know this in order to have 
sufficient information for applying the ABL rule~(\ref{ABL}) with the trivial property $\bf 1$ in place of 
$\ketbra bb$.] For predictions, we must use the Born rule. The Born rule, on the other hand, is of little 
use when it comes to sounding the ontological implications of quantum mechanics, for these must be 
based on objective probabilities (which are free of any element of ignorance), and such probabilities, as was shown in Sec.~2, are ABL probabilities.
 
With Mermin~\cite{Mermin98} I believe that all the mysteries of quantum mechanics can be reduced 
to the single puzzle posed by the existence of objective probabilities. As I see it, the key to this 
puzzle is the contingent reality of spatial and temporal 
distinctions~\cite{Mohrhoff00}. We are neurophysiologically disposed to 
think of space as something that exists by itself (rather than by virtue of the relational properties of 
matter), that contains matter (rather than spatial relations between material objects), and that is 
intrinsically and infinitely differentiated and thus adequately represented by a set of points that 
can be labeled by triplets of real numbers~\cite{MohrhoffCCP}. Yet if all conceivable 
spatial divisions were intrinsic to space, they would have an unconditional reality, and one of the 
following statements would necessarily be true of every object $S$ contained in the union $ R\cup R'$ 
of two spatial regions: (i)~$S$ is inside $R$; (ii)~$S$ is inside $R'$; (iii)~$S$ has two parts, one in $R$ and one in $R'$. No particle 
could ever pass through the union of two slits without passing through either slit in 
particular {\it and} without consisting of parts that pass through different slits. But this is precisely what 
particles do when interference fringes are observed in two-slit experiments (and we do not postulate 
hidden variables). Hence, spatial divisions cannot be intrinsic to space. They have a contingent 
reality. Like the contingent properties of quantum-mechanical systems, they are extrinsic; they 
supervene on the actual goings-on in the physical world, and they may be real for one object and 
nonexistent for another. 
 
It is a fundamental principle of quantum mechanics that the probability of a process $\cal P$ capable 
of following several alternatives depends on whether or not the alternative taken by the process is 
indicated or capable of being indicated. If something indicates the alternative taken or if this is 
capable of being indicated, the probability of the process is given by the sum of the probabilities 
associated with its alternatives. By ``capable of being indicated'' I mean that the alternatives of $\cal 
P$ are correlated with the alternatives of another process ${\cal P}'$ such that a determination of the 
alternative taken by ${\cal P}'$ reveals the alternative taken by $\cal P$. (Paradigm examples of this 
kind of a situation are the experiments of Einstein, Podolsky, and Rosen~\cite{EPR,Bohm51} and of 
Englert, Scully, and Walther~\cite{ESW91,ESW94,Mohrhoff99}.) On the other hand, if nothing either 
indicates or is capable of indicating the alternative taken by the process, the probability of the process 
is given by the absolute square of the sum of the amplitudes associated with the alternatives. What 
is the meaning of this fundamental principle? 
 
I submit that if the alternative taken is indicated, we add probabilities {\it because} in this case the 
conceptual distinction that we make between the alternatives has a reality for the process or the 
system undergoing it---the distinction corresponds to something in the objective world. If the 
alternative taken is neither indicated nor capable of being indicated, we add amplitudes {\it because} 
in this case the conceptual distinction that we make between the alternatives has no reality for the 
process or the system undergoing it---the distinction corresponds to nothing in the objective 
world~\cite{noteQA}. In our three-hole experiment it is the presence of $F_A$ that makes the spatial 
distinction between $A$ and $B\cup C$ a reality for the particle. Assuming the beepers to be 100\% 
efficient~\cite{noteEU}, the presence of $F_A$ warrants two objective truth values, one for the 
proposition ``The particle goes through $A$'' and one for the proposition ``The particle goes through 
$B\cup C$''. The existence of these two truth values warrants the reality, for the particle, of the 
distinction between the two spatial regions $A$ and $B\cup C$. The reality of this distinction is the 
reason why in the denominator of the right-hand side of eq.~(\ref{boxes}) we add the 
probabilities associated with the alternatives represented by ${\bf P}_A$ and ${\bf P}_{B\cup C}={\bf 
P}_B+{\bf P}_C$ (eq.~\ref{deno1}). By the same token, the presence of both beepers warrants objective truth values for three 
propositions, ``The particle goes through $A$'', ``The particle goes through $B$'', and ``The particle 
goes through $C$''. The existence of these three truth value warrants the reality, for the particle, of 
the distinction between the three spatial regions $A$, $B$ and $C$. The reality of this distinction 
is the reason why in the denominator of the right-hand side of eq.~(\ref{boxes}) we add the 
probabilities associated with the alternatives represented by ${\bf P}_A$, ${\bf P}_B$, and ${\bf P}_C$ 
(eq.~\ref{deno2}). 
 
It needs to be stressed that this explanation of the uncertainty principle---as stated by Feynman and 
Hibbs~\cite{FH65}: Any determination of the alternative taken by a process capable of following more 
than one alternative destroys the interference between alternatives---owes nothing to the ABL rule. It 
rests on the fact that a successful measurement of an observable $Q$ with $n$  eigenvalues warrants 
attributing $n$ truth values to $n$ propositions, and thus warrants the objective  distinctness of the 
$n$ alternatives. The contextuality of ABL probabilities (as well as Born 
probabilities~\cite{noteCT}) merely emphasizes the contingent reality of the distinctions we are wont 
to make. Given  two observables with a common eigenvalue $q$, the ABL probability associated with 
$q$ in general depends on which of the two observables is being measured---it depends on the entire 
spectrum of the observable being measured. In particular, the probability of finding that the particle 
goes through $A$ depends on  whether or not our distinction between $B$ and $C$ is objectively 
warranted (that is, whether or not  the measurement is capable of distinguishing between ``The 
particle goes through $B$'' and ``The particle goes through $C$''). 
 
The ontological significance of objective probabilities, then, is that they signal the unreality of some of the conceptual distinctions that we make. 
Since the warranted distinctions depend on what precisely is indicated, so do the probabilities of the corresponding alternatives.

If we conceptually partition space into smaller and smaller regions, we eventually arrive at a partition 
$\{R_i\}$ into {\it finite} (rather than infinitesimal) regions that are so small that the distinctions we make between them have no 
reality at all~\cite{Mohrhoff00,MohrhoffCCP}. Our spatial distinctions bottom out in a sea of objective 
probabilities. At a scale at which position-indicators (``detectors'') with sufficiently small and sufficiently localized 
sensitive regions no longer exist, all we can say is counterfactual and probabilistic. This tells us that 
the  world is only finitely differentiated spacewise. Conversely, the limited spatial differentiation of the  
objective world finds its proper expression in counterfactual assignments of objective probabilities. 
 
What is true of the world's spatial aspect is equally true of its temporal aspect. There is no such thing 
as an intrinsically differentiated time, and therefore not only the contingent properties of things but 
also the times at which they are possessed are extrinsic. What is temporally differentiated is physical 
systems, and every physical system is temporally differentiated to the extent that it passes through 
successive states, in the proper sense of ``state'' that connotes properties indicated by facts. And 
since no finite system passes through an infinite number of successive states in a finite time span, no 
such system is infinitely differentiated timewise. The times that exist for a system $S$ are the 
(factually warranted) times at which it has (factually warranted) 
properties~\cite{Mohrhoff00,MohrhoffCCP}. 
 
How, then, are we to conceive of system $S$ during the interval between the times $t_a$ and $t_b$? 
Since during this interval $S$ lacks factually warranted properties, all that can be said about $S$ 
between $t_a$ and $t_b$ is counterfactual and probabilistic. Our conceptual temporal distinctions, 
too, bottom out in a sea of objective probabilities. Not only is there no state (in the proper sense just 
defined) that obtains during this interval, but also there is no time between $t_a$ and $t_b$ at which 
any state could obtain. 
 
The importance of this result cannot be overemphasized. It entails that the parameter $t$ 
appearing in the (prior) Born probability $\absosq{\braket{a(t)}{q_i}}$, in the posterior Born probability 
$\absosq{\braket{q_i}{b(t)}}$, and in the ABL probability 
$\absosq{\braket{a(t)}{q_i}\braket{q_i}{b(t)}}/\sum_j\absosq{\braket{a(t)}{q_j}
\braket{q_j}{b(t)}}$ 
cannot be interpreted as the time at which anything obtains {\it per se}. The parameter $t$ refers to 
the time of a measurement. Only if a measurement is actually made does it represent a time that 
exists for $S$ because only then is there a contingent property that can be attributed to $S$ at the 
time $t$. If the measurement is not actually made, the time at which it is made in a possible world 
does not exist for the actual-world edition of $S$. It follows in particular that neither the prior 
probability measure $\ket{a(t)}$ nor the posterior probability measure $\ket{b(t)}$ nor the ABL 
probability measure $\ts{a(t)}{b(t)}$~\cite{noteTS} can be interpreted as something that obtains at the 
time $t$~\cite{notePM}. If the measurement is made, what obtains at time $t$ is a result rather than a 
probability measure, and if the measurement is not actually made, there is no time $t$ at which 
anything concerning $S$ could obtain. 
 
\section{\large QUANTUM COUNTERFACTUALS\\ 
AND THE ``FLOW'' OF TIME} 
 
In this section I want to point out a common but unwarranted assumption about the temporal aspect of 
the physical world, and I want to show that this assumption leads to the views that Kastner is 
defending in her Comment~\cite{Kastner}. Needless to say, I cannot avow 
that she actually makes this assumption. 
 
The conclusions we have reached in the previous section run counter to a common way of thinking 
about time, according to which the experiential now and the temporal distinctions that we base on it 
are features of the physical world---the world accessible to physics. The experiential now is temporally unextended and 
undifferentiated. If it did correspond to something in the physical world, this would seem to warrant the notion of an objective {\it 
instantaneous} state that evolves in an infinitely differentiated time. Yet this contradicts the fact 
that the world is only finitely differentiated timewise. 
 
In truth, nothing in the physical world 
corresponds to the experiential now and the temporal distinctions that we base on it. There simply is 
no objective way to characterize the present or to distinguish between the past, the present, and the 
future. These distinctions can be characterized only subjectively, by how they relate to us: 
through memory, through the present-tense immediacy of qualia (introspectible properties like pink or 
turquoise), or through anticipation. In the world accessible to physics we may qualify events or states 
of affairs as past, present, or future {\it relative to} other events or states of affairs, but we cannot 
speak of {\it the} past, {\it the} present, or {\it the} future. 
 
In classical physics this is not blindingly obvious. Classical physics is consistent with the notion of an 
instantaneous state that evolves in an infinitely differentiated time, and that encapsulates not only 
possessed properties but also everything that (i)~happened or obtained at earlier times and (ii)~is 
causally relevant to what happens or obtains at later times. This is how we come to conceive of 
``fields of force'' that evolve in time (and therefore, in a relativistic world, according to the principle of 
local causality) and that causally link earlier times to later times (and therefore, in a relativistic world, 
local causes to their distant effects). But this does not entail that the notion of an {\it evolving} 
instantaneous state is itself consistent. If we conceive of temporal relations in the physical world, we 
conceive of their relata {\it at the same time} even though they happen or obtain {\it at different 
times}. Since we can't help it, that has to be OK. But it is definitely not OK if we introduce into our 
simultaneous and spatial mental picture of a temporal whole anything that evolves or advances 
across this temporal whole. One cannot represent a spatiotemporal whole as a simultaneous spatial 
whole and then imagine the present as advancing through it or an instantaneous state as evolving in 
it. To do this is to depict the spatiotemporal whole $\cal U$ as persisting unchanged in a time that is 
extraneous to $\cal U$, and to depict something as advancing or evolving across the unchanging 
$\cal U$ in that extraneous time. There is only one time, the fourth dimension of space-time. There is 
not another time in which anything evolves or advances across space-time as if space-time 
itself---rather than our mental picture of it---were a persisting and unchanging whole. If the present is 
anywhere in the spatiotemporal whole, it is trivially and vacuously everywhere---or, rather, 
everywhen. 
 
To philosophers the perplexities and absurdities entailed by the notion of an advancing present or a 
flowing time are well known~\cite{time}. Physicists began to recognize the subjectivity of the present 
and the nonexistence of an {\it evolving} instantaneous state with the discovery of the relativity of 
simultaneity. In the well-known words of Hermann Weyl, ``The objective world simply {\it is}; it does 
not {\it happen}. Only to the gaze of my consciousness, crawling upward along the life line of my 
body, does a section of this world come to life as a fleeting image in space which continuously 
changes in time.''~\cite{Weyl49} Yet the non-objectivity of the now remains deeply 
counterintuitive~\cite{noteCI}. Where space is concerned, we have no difficulty in abstracting from 
our subjective, perspectival point of view and adopt ``the view from nowhere''~\cite{Nagel86}. Where 
time is concerned, we find it incomparably more difficult to abstract from our subjective, 
present-centered point of view and adopt ``the view from nowhen''~\cite{Price96}, not least because 
this seems to conflict with our incorrigible self-perception as free agents, which seems to require an 
open future~\cite{noteOF}. 
 
In the world of physics, the future is as closed as the past. (This follows directly from the 
non-objectivity of the distinction between the future and the past.) If system $S$ has property $\ketbra 
aa$ at time $t_a$ and property $\ketbra bb$ at time $t_b$, then it always has been and always will be 
true that system $S$ has these properties at the respective times $t_a$ and $t_b$. There is nothing 
objectively or physically open about this. All that is ``open'' in respect of this is our knowledge, prior to 
these times, of the properties possessed at these times. What {\it is} (and always has been and 
always will be) objectively or physically open is the results of unperformed measurements. That is 
why we can assign to them objective probabilities. 
 
If we retain the fallacious notion of the objectivity of the now and of the temporal distinctions that are 
based on it, or if we subscribe to the ensuing idea of an evolving instantaneous state, we are 
committed to regarding system $S$ as infinitely differentiated timewise and to attributing to $S$ a 
state that obtains at every instant of time. Quantum mechanics offers us not one but three candidates 
for such a state: the ``prepared'' or ``retarded'' state $\ket{a(t)}$, the ``retropared'' or ``advanced'' state 
$\ket{b(t)}$, and the time-symmetric two-state $\ts{a(t)}{b(t)}$. For reasons that are psychological 
rather than physical, quantum realists usually settle for $\ket{a(t)}$, which leads to the well-known 
measurement problem. (The projection postulate is as mysterious when applied to evolving states of 
affairs as it is trivial when applied to probability measures.) 
 
Another consequence of the myth of an evolving instantaneous state is that a difference in what 
obtains at the intermediate time $t$ must make a difference to what obtains at later (earlier) times 
given that it makes no difference to what obtains at earlier (later) times. The intervening 
measurement ``disturbs the system''---an ubiquitous but illegitimate phraseology found, for instance, 
in Sharp and Shanks~\cite{SS93},---with the result that the ``disturbed'' system is necessarily 
different from the ``undisturbed'' system either before or after the ``disturbance'' (or both). This notion 
implies an apparent infringement of cotenability: The state that obtains before the hypothetical 
measurement of $Q$ and the state that obtains after this measurement cannot be both independent 
of the truth value of the antecedent of~(A). As Kastner puts it, ``holding fixed {\it both} pre- and post-selection 
states''~\cite{Kastner} is impossible. If the prepared state $\ket a$ obtains before the $Q$ measurement at time $t$ then 
either $\ket a$ or one of the eigenstates of $Q$ obtains after the time $t$, depending on whether or 
not the measurement is made. If the retropared state $\ket b$ obtains after the time $t$ then either $\ket b$ or one of the 
eigenstates of $Q$ obtains before the time $t$, depending on whether or not the measurement is 
made~\cite{noteTF}. (If one thinks of $\ts ab$ as the state that obtains between $t_a$ and $t_b$ in 
the absence of an intervening measurement, an intervening measurement changes both the earlier 
and the later state: If the measurement yields $q_i$, the former state becomes $\ts a{q_i}$ while the 
latter state becomes $\ts{q_i}b$.) 
 
The relevant background conditions, however, are not probability measures, nor are these states that obtain. The relevant background conditions are the possessed 
properties that determine probability measures. The unmeasured system differs from its 
possible-world counterpart neither at the times $t_a$ and $t_b$ (at which times the same properties 
are possessed in both worlds) nor during the intervals between $t_a$ and $t$ and between $t$ and 
$t_b$ (during which intervals no properties are possessed in either world) but only at the time $t$, 
which exists for the measured system but not for the unmeasured one. The intervening measurement 
has no influence whatsoever on what obtains at any other time. It has an influence on some 
probability measures but none on the {\it relevant} probability measures. The relevant probability 
measures are the prior probability measure $\ket a$ for times {\it earlier} than $t$ and the posterior 
probability measure $\ket b$ for times {\it later} than $t$, while what is affected by the intervening 
measurement is the prior probability measure $\ket a$ for times {\it later} than $t$ and the posterior 
probability measure $\ket b$ for times {\it earlier} than $t$. The ABL probability for the time $t$ is 
obtained by combining, according to Bayes' theorem, the probabilities that are unaffected by the 
measurement, and thus it is independent of whether or not the measurement is actually performed. 
 
Cohen~\cite{Cohen95} appears to reject the counterfactual use of the ABL rule not only on the basis 
of the intrinsically inconsistent eq.~(\ref{BS}) but also on the ground that that use is ``not 
consistent with realist interpretations of quantum mechanics.'' This, however, is no ground for 
rejecting counterfactual uses of the ABL rule. Rather, it is ground for rejecting realist interpretations of 
quantum mechanics. By ``realist interpretations'' Cohen may mean interpretations that endorse either 
Redhead's sufficiency condition~\cite{Redhead87}, according to which Born probability one implies 
the existence of an element of reality, or the so-called ``eigenstate-eigenvalue link,'' according to 
which ``being in'' an eigenstate of some observable implies the same. In point of fact, neither does imply the 
existence of an element of reality~\cite{Mohrhoff00}. Or else Cohen may mean interpretations that 
construe probability measures as instantaneous states that evolve in an infinitely differentiated time. 
Such interpretations involve the double error of (i)~treating time-dependent probability measures as if 
they were actual states and of (ii)~extending temporal distinctions beyond the limits within which they are 
objectively warranted. Is instrumentalism the sole alternative to such realist interpretations? I submit 
that, on the contrary, the correct {\it ontological} interpretation of quantum mechanics can be found only when such realist 
interpretations are rejected. The highlights of that interpretation are the contingent reality of spatial 
and temporal distinctions and the existence of finite limits to the spatial and temporal differentiation of 
the physical world. 
 
\vspace{\baselineskip}{\bf\noindent Acknowledgment} 
 
\noindent I wish to thank Ruth Kastner for a stimulating exchange of views that was of considerable 
help in preparing this article.


\begin{thebibliography}{99}
\bibitem{noteCP} 
A contingent property of system $S$ is a property that $S$ may but does not necessarily (that is, by 
definition) possess. It corresponds to an eigenvalue of a quantum-mechanical observable on $S$.
\bibitem{notePA} 
If possessed properties are attributable, properties capable of being possessed are ``potentially 
attributable.''
\bibitem{note1} 
Much confusion can be avoided by maintaining a clear distinction between projection operators 
representing probability measures and projection operators representing potentially attributable 
properties. If a system with a vanishing Hamiltonian is observed to possess the property 
$\ketbra\psi\psi$ at the time $t_0$, the Born probabilities of the possible results of any measurement 
performed at a time $t>t_0$ are given by the probability measure $\ketbra\psi\psi$. Sloppiness about 
the distinction between probability measures and properties leads to the insidious notion that the 
system possesses the property $\ketbra\psi\psi$ not only at the time $t_0$, when it is factually 
warranted, but also at the time $t>t_0$, when the possession of this property is not warranted by any 
facts~\cite{Mohrhoff00}.
\bibitem{noteB} 
The quotation signs serve as reminder that all that is actually prepared is a probability measure.
\bibitem{noteIP} 
The converse is not true: If $\braket ab=0$, obtaining $\ketbra aa$ at $t_a$ and $\ketbra bb$ at $t_b$ 
is nomologically impossible unless another measurement (of an observable that has neither $\ket a$ 
nor $\ket b$ among its eigenkets) is interposed.
\bibitem{notePP} 
Kastner uses the term ``posterior probability'' in a different sense, namely for the trivial probabilities 
(either zero or one) that are associated with the possible results of an actually performed 
measurement when the actual result is taken into account.
\bibitem{noteQA} 
There remains the question of why we add probabilities if the alternative taken is not indicated but 
capable of being indicated. If a process is capable of following one set $\{{\cal A}^1_k\}$ of mutually 
contradictory alternatives, it is in principle capable of following at least one other set $\{{\cal A}^2_k\}$ 
of mutually contradictory alternatives such that the two sets are complementary. In the experiment of 
Englert, Scully, and Walther, for instance, experimenters may determine either the slit taken by an 
atom (by identifying the microwave cavity containing the photon that is left behind by the atom) or 
the phase relation with which the atom emerges from the slit plate (by opening the electro-optic 
shutters and watching the photosensor situated between the cavities). If the experimenters determine 
neither the slit taken nor the phase relation, then not only the conceptual distinctions we make 
between the alternatives of each set correspond to nothing in the objective world, but also the 
conceptual distinction we make between the two sets has no counterpart in the objective world. In this 
case we cannot add the amplitudes associated with either set of alternatives, for the amplitudes 
associated with one set are undefined unless the process takes a particular alternative from the 
complementary set. In particular, the amplitudes associated with the first set of alternatives (the atom 
takes this or that slit) are undefined unless the atom takes both slits with a particular phase relation.
\bibitem{noteEU} 
Even when we measure prior Born probabilities, we postselect: We discard measurements that did 
not succeed. Our euphemism for this procedure is that we use apparatuses that are 100\% efficient. 
Such apparatuses are physically unrealizable on principle~\cite{Mohrhoff00}.
\bibitem{notePM} 
This ought to be self-evident. The probability of finding the particle in box $A$ at time $t$ is not 
something that obtains at the time $t$, anymore than it is something that exists inside box $A$.
\bibitem{noteCI} 
In the past I myself have, in increasingly moderate terms, argued for the objectivity of the temporal 
modes past, present, and future. I now consider it more fruitful if in our attempts to understand the 
larger world that includes consciousness we face up squarely to the subjectivity of the temporal 
modes.
\bibitem{noteOF} 
This conflict is spurious. As was pointed out in Ref. \cite{Mohrhoff00}, even if the future ``already'' 
existed in some way, my present choice can be one of its determining factors. What would be 
inconsistent with my self-perception as a free agent would be the possibility of knowing the future 
(and hence of knowing my choice before it was made).
\end{thebibliography}
\end{document}